\documentclass[12pt]{article}
\usepackage{graphicx}
\usepackage{xspace}

\newcommand\pubnumber{}
\newcommand\pubdate{\today}

\def\warwick{Physics Department \\
University of Warwick, United Kingdom}

\def\onbehalfcdf{\footnote{On behalf of the CDF Collaboration}}

\def\Title#1{\begin{center} {\Large #1 } \end{center}}
\def\Author#1{\begin{center}{ \sc #1} \end{center}}
\def\Address#1{\begin{center}{ \it #1} \end{center}}

\newcommand\pubblock{\rightline{\begin{tabular}{l} \pubnumber\\
         \pubdate  \end{tabular}}}
\newenvironment{Abstract}{\begin{quotation}  }{\end{quotation}}
\newenvironment{Presented}{\begin{quotation} \begin{center} 
             PRESENTED AT\end{center}\bigskip 
      \begin{center}\begin{large}}{\end{large}\end{center} \end{quotation}}
\def\Acknowledgements{\bigskip  \bigskip \begin{center} \begin{large}
             \bf ACKNOWLEDGEMENTS \end{large}\end{center}}
\newcommand{\niceMKbabar}{\mbox{\sl B\hspace{-0.4em}
{\small\sl%
A}\hspace{-0.37em} \sl B\hspace{-0.4em} {\small\sl%
A\hspace{-0.02em}R}}}

\newcommand{\Jpsi}{\ensuremath{J\!/\!\psi}\xspace}
\newcommand{\invfb}{\ensuremath{\mathrm{fb^{-1}}}\xspace}
\newcommand{\BsJpsiphi}{\ensuremath{\Bs\rightarrow\Jpsi\phi}\xspace}
\newcommand{\Bs}{\ensuremath{B_s}\xspace}

\begin{document}
\begin{titlepage}
\pubblock

\vfill
\Title{Measurement of $B_s$ mixing phase at CDF.}
\vfill
\Author{ Michal Kreps\onbehalfcdf}
\Address{\warwick}
\vfill
\begin{Abstract}
We present improved bounds on the $CP$-violating phase
$\beta_s^{\Jpsi\phi}$ and on the
decay-width difference $\Delta\Gamma$ of the neutral
$B_s^0$ meson system obtained by the CDF experiment at the
Tevatron collider . We use 6500 $B_s^0 \rightarrow J/\psi
\phi$ decays collected by the dimuon trigger and reconstructed 
in a sample corresponding to integrated luminosity of 5.2 fb$^{-1}$.
Besides exploiting a two-fold increase in statistics with
respect to the previous measurement, several improvements
have been introduced in the analysis including
a fully data-driven flavour-tagging calibration and proper
treatment of possible S-wave contributions.
\end{Abstract}
\vfill
\begin{Presented}
6th International Workshop on the CKM Unitarity Triangle \\
University of Warwick, United Kingdom, September 6-10, 2010 
\end{Presented}
\vfill
\end{titlepage}
\def\thefootnote{\fnsymbol{footnote}}
\setcounter{footnote}{0}

\section{Introduction}

Since the discovery of $CP$ in 1964 in neutral kaon system
\cite{Christenson:1964fg}, $CP$
violation plays crucial role in the development of standard
model (SM) and probing a new physics (NP). In 1973 Kobayashi and Maskawa
proposed, as one of the possible explanations for $CP$
violation in kaon system, extension to six quarks model
\cite{Kobayashi:1973fv} in which the $CP$
violation is explained through the quark mixing parametrised by
the Cabibbo-Kobayashi-Maskawa (CKM) matrix.
A single complex phase in the CKM matrix is responsible for
all $CP$ violation in the SM.
The observation of large $CP$ violation
in $B^0$ mesons by \niceMKbabar{} and Belle experiments
\cite{BelleBabar} confirmed the SM and paved way to Nobel
Prize award to Kobayashi and Maskawa in 2008.

After confirmation of the SM focus shifted to
search for a NP. One of the most
promising processes is $B_s$ mixing governed by the CKM
matrix element $V_{ts}$.  
The indirect information defines 
$V_{ts}$ to be almost real, which translates to the fact that
the $CP$ violation due to the $B_s$ 
mixing is expected to be tiny in the SM. First measurements performed
by the CDF and D\O{} experiments \cite{CDFD0} showed about 1.5-2.0 $\sigma$
discrepancy with the SM, which caused large
excitement in the community. In this proceedings we review
most important aspects and results of the updated CDF
analysis, using dataset corresponding to integrated luminosity
of $5.2$ $\mathrm{fb^{-1}}$ \cite{cdf:betas}.

\vspace*{-0.15cm}
\section{Candidate selection}

To select candidates we use an artificial neural network (ANN)
trained on a signal sample from simulation and data events
from a $B_s$ mass sidebands as background. Inputs to the
ANN are transverse momenta of $B_s$ and $\phi$
mesons, particle identification for kaons and
muons and quality of the kinematical fits to the candidates.
For the result presented here we choose requirement on the
ANN output which minimises
uncertainties on the $CP$ violating phase
$\beta_s^{\Jpsi\phi}$. The best point is found by performing
simulated experiments with three different true values of
$\beta_s^{\Jpsi\phi}$ (0.02, 0.3 and 0.5) and single value
for $\Delta\Gamma$ and amplitudes 
 with amount of signal and background
corresponding to different requirements on the 
ANN output and selecting the value which provides smallest
parabolic uncertainty on the $\beta_s^{\Jpsi\phi}$. 
%
\begin{figure}[htb]
\centering
\includegraphics[width=5.5cm]{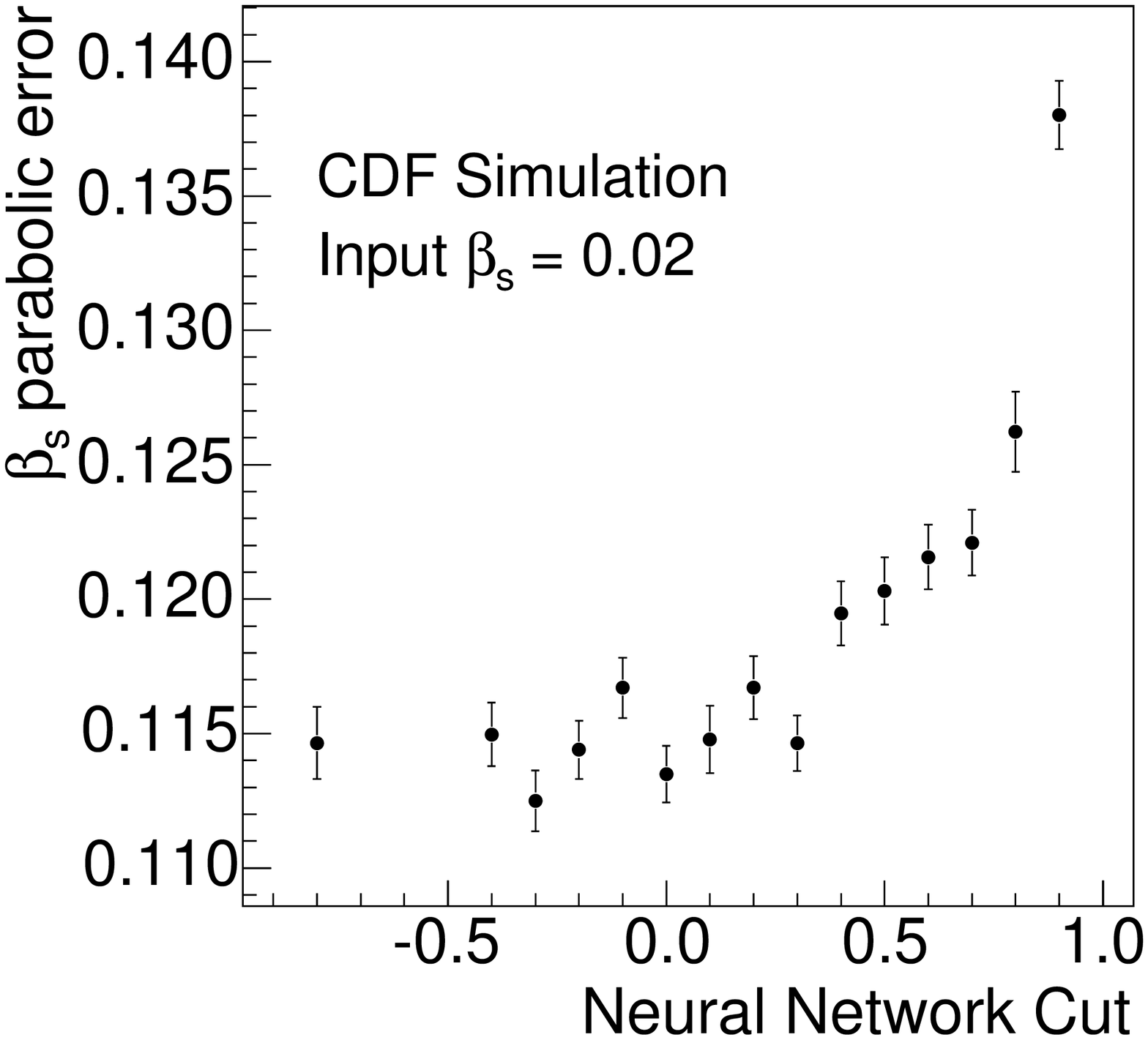}
\includegraphics[width=7cm]{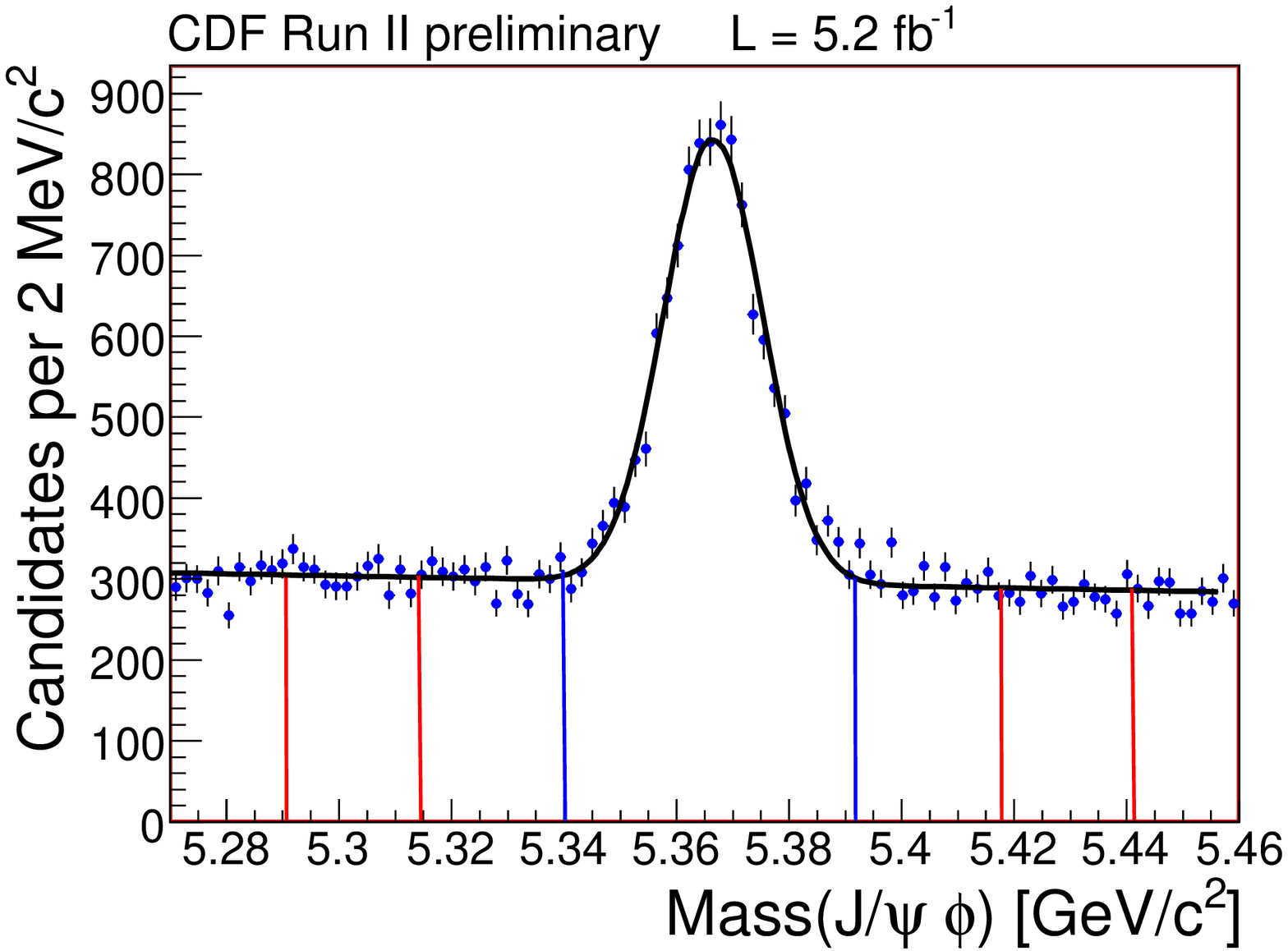}
\caption{An example of the dependence of parabolic uncertainty
of $\beta_s^{\Jpsi\phi}$ on ANN output
requirement (left). The invariant mass distribution of the
selected candidates used in the analysis (right).}
\label{fig:selection}
\end{figure}
In Fig.~\ref{fig:selection} we show an example of such
simulation for true $\beta_s^{\Jpsi\phi}$ value at the SM 
 expectation. As result we select candidates with
the ANN output larger than 0.2, which is less
stringent requirement than one used in previously.
The resulting invariant mass distribution contains about
6500 signal events and is shown in Fig.~\ref{fig:selection}. 

\vspace*{-0.15cm}
\section{Flavour tagging}

The flavour tagging is one of the important tools in the
analysis. Its task is to determine whether reconstructed
candidate was produced as $B_s$ or $\overline{B}_s$. 
Two algorithms are used at hadron colliders. The first one,
called opposite side tagging, explores the fact that $b$-quarks
are dominantly produced in $b\overline{b}$ pairs, so
determination of the flavour of other $b$-hadron determines
also the flavour of signal one. The second algorithm, called same
side tagging, explores properties of hadrons produced in
hadronization of $b$ quark into $B_s$ mesons. 

The opposite side algorithm explores several sources of
information 
from the non-signal $b$-hadron in an
event. The most clean explores the fact that
about 10\% of $b$-hadrons decay to a final state containing
lepton. At CDF we use electrons and muons and given that
most of them arise from the $b$-hadron decays in events with
reconstructed $B_s$, they provide clean
information, but at expanse of small efficiency. The second
source  uses fact the most abundant
$b\rightarrow c\rightarrow s$ quark level transition
yielding a final
state which often contains charged kaon, which determines
the flavour
of the $b$-hadron. The most efficient source identifies
jet containing $b$ quark and calculates weighted
charge of the tracks in jet to determine, whether original
quark had positive or negative charge and thus determine
flavour. While this algorithm is efficient, its purity is
rather small. We combine all three sources into
single decision using neural network. The quality of decision is predicted for
each event and checked on the fully reconstructed $B^+$ events.
The overall performance of the algorithm is about 1.2\%, which
can be understood as having 1.2\% of the overall statistics with
perfectly known production flavour.

The same side algorithm exploits fragmentation process. 
To form a $B_s$ out of $\overline{b}$ quark we need to
attach $s$ quark to it. The $s$ quark comes from a pair
generated out of the vacuum and after forming $B_s$ an $\overline{s}$ quark
remains to form other hadron. If it ends up in the charged kaon
it can be used to determine the production flavour.
\begin{figure}[tbh]
\centering
\includegraphics[width=5.2cm]{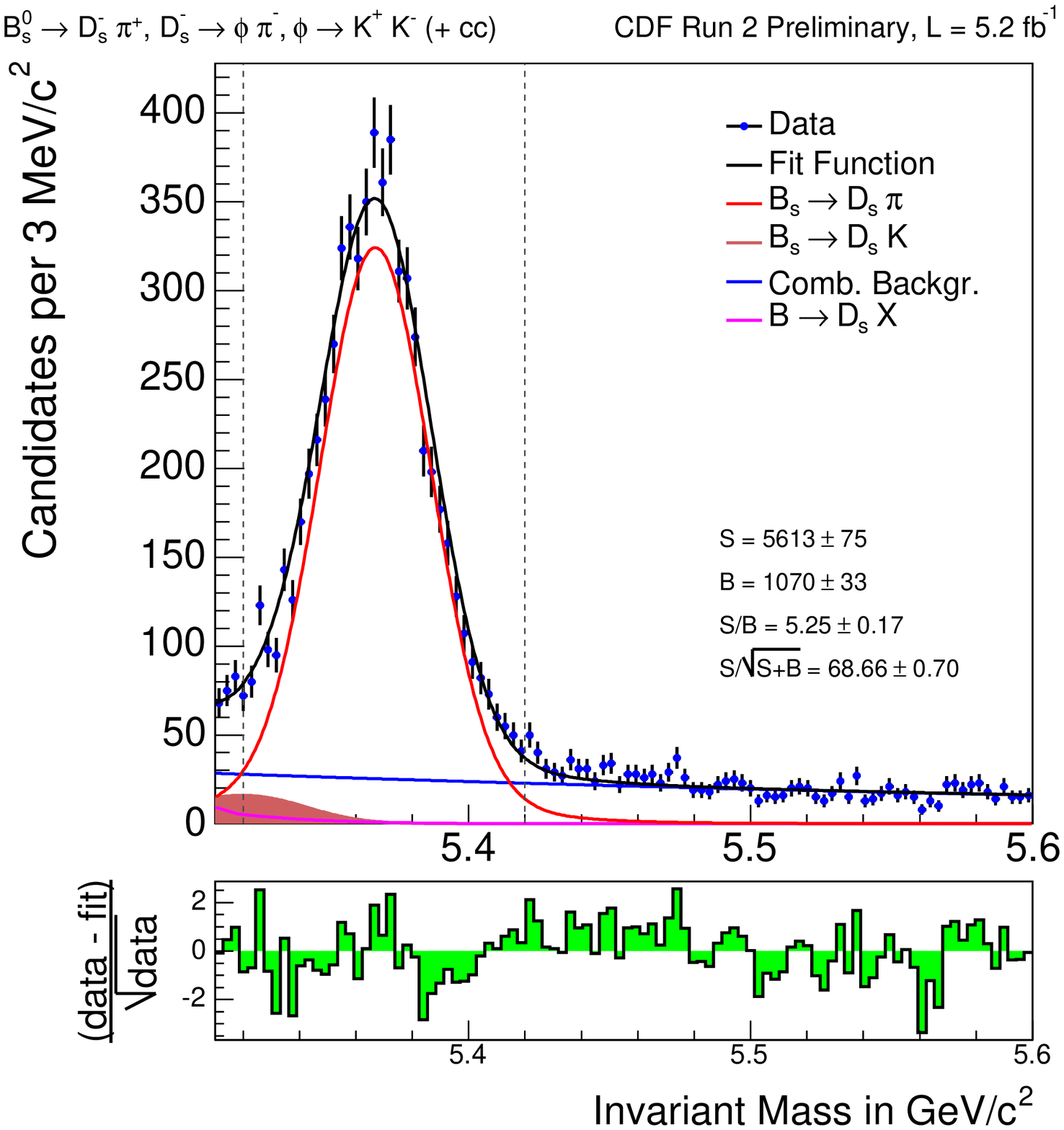}
\includegraphics[width=5.8cm]{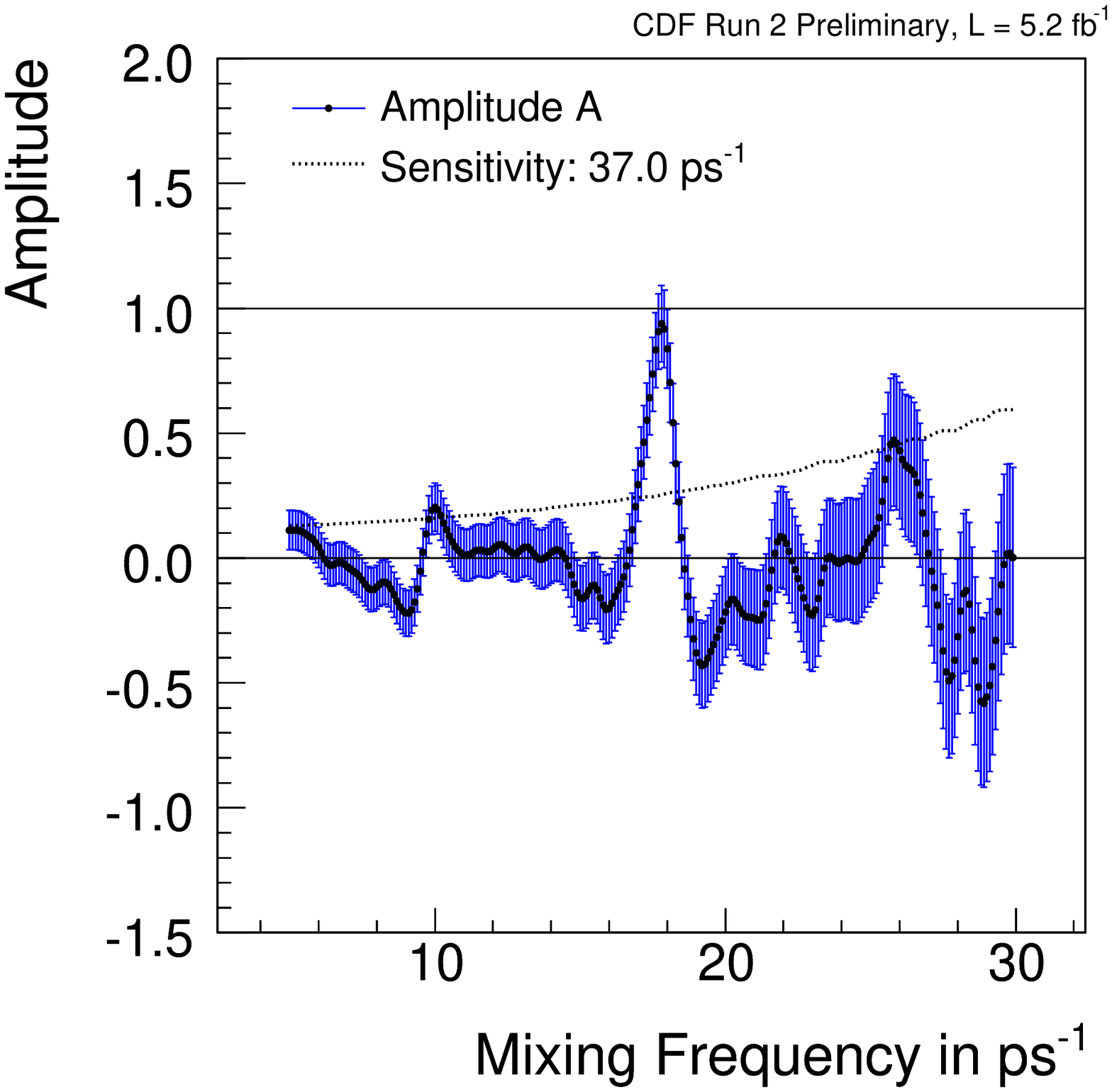}
\caption{The invariant mass distribution of $B_s\rightarrow
D_s\pi$ with $D_s\rightarrow \phi\pi$ (left) and the
amplitude scan in the full dataset (right).}
\label{fig:SSKT}
\end{figure}
The challenge is to identify right kaon in an event with many
tracks. Another challenging part is to calibrate quality of
the decisions on a data as only process available is $B_s$
mixing. Despite all challenges, CDF uses this
algorithm with good success. In calibration we use data
spanning same period as our dimuon dataset. 
For calibration we reconstruct
about 12900 $B_s\rightarrow D_s(3)\pi$ signal events. The
predicted quality of decision for each event is scaled by
a global scaling factor which is determined in the fit for
$B_s$ oscillations. 
We show the invariant mass distribution of channel contributing
about half of the statistics together with the amplitude scan on
full sample in Fig.~\ref{fig:SSKT}. The resulting tagging performance is about
3.2\%. The obtained $B_s$ mixing frequency is $\Delta
m_s=17.79\pm0.07$ ps$^{-1}$ with uncertainty being statistical only.
Full details of this calibration can be found in
Ref.~\cite{cdf:betas}.

\vspace*{-0.1cm}
\section{Fit description}

On the limited space available we cannot describe all
details of the fit, so we will touch
only main features with some emphasis on changes compared to
the previous versions of the analysis. The full details are spelled out
in Ref.~\cite{Azfar:2010nz}. As spin zero $B_s$ decays into
two spin one particles (\Jpsi and $\phi$) three different
amplitudes corresponding to different angular momentum are
involved. In CDF we use a basis in which three amplitudes are
written in terms of polarisation amplitudes. The three
amplitudes give six angular terms, three terms
corresponding to the squares of  amplitudes and three
corresponding to the interference between amplitudes. Each of
the six terms has its own angular and decay time dependence.
Some of the terms exhibit usual $\sin(\Delta m_s t)$
behaviour and some are proportional to $\cos(\Delta m_s t)$.
Thanks to the non-zero decay width difference ($\Delta\Gamma$) in the $B_s$
system, depending on the size of the $\Delta\Gamma$ 
and polarisation amplitudes one can gain
considerable information on the $CP$ violating phase also
without resolving the oscillation pattern.

Since the first analysis there is an ongoing discussion whether
there can be contribution from non-resonant $\Jpsi K^+K^-$
or $\Jpsi f_0(980)$ with $f_0(980)\rightarrow K^+K^-$
decays, collectively named s-wave. Those were neglected in
previously and there is an open question about the size of possible
contribution and whether it can bias result. In the
presented result we introduce an s-wave component which is
allowed to float during the analysis. This additional
component introduces four new terms, one corresponding to
the square of the s-wave amplitude and three for the interference
between s-wave contribution and decay to $\Jpsi\phi$. 

\section{Results}

\begin{figure}
  \centering
  \includegraphics[width=5cm]{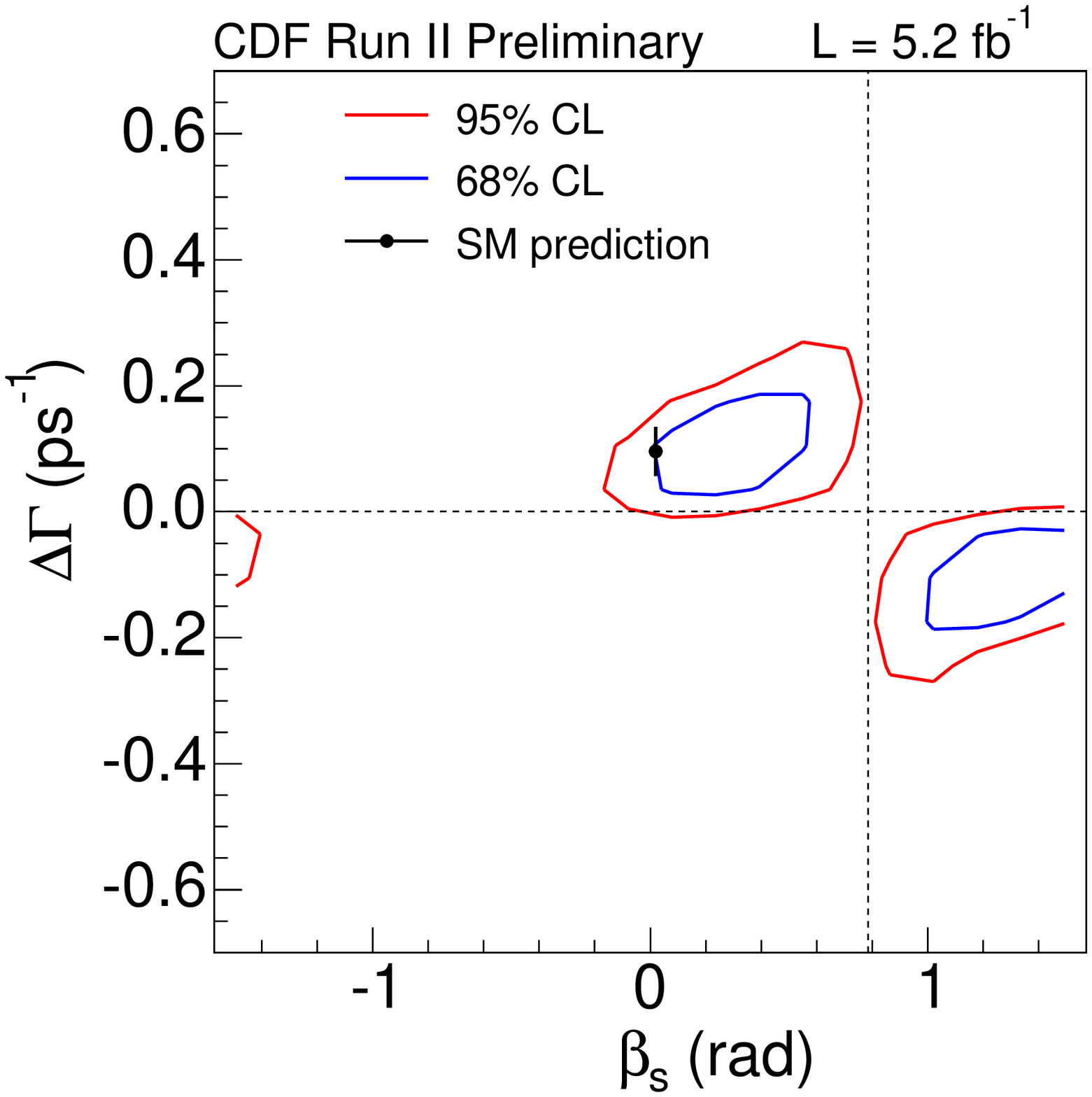}\hspace{0.5cm}
  \includegraphics[width=5cm]{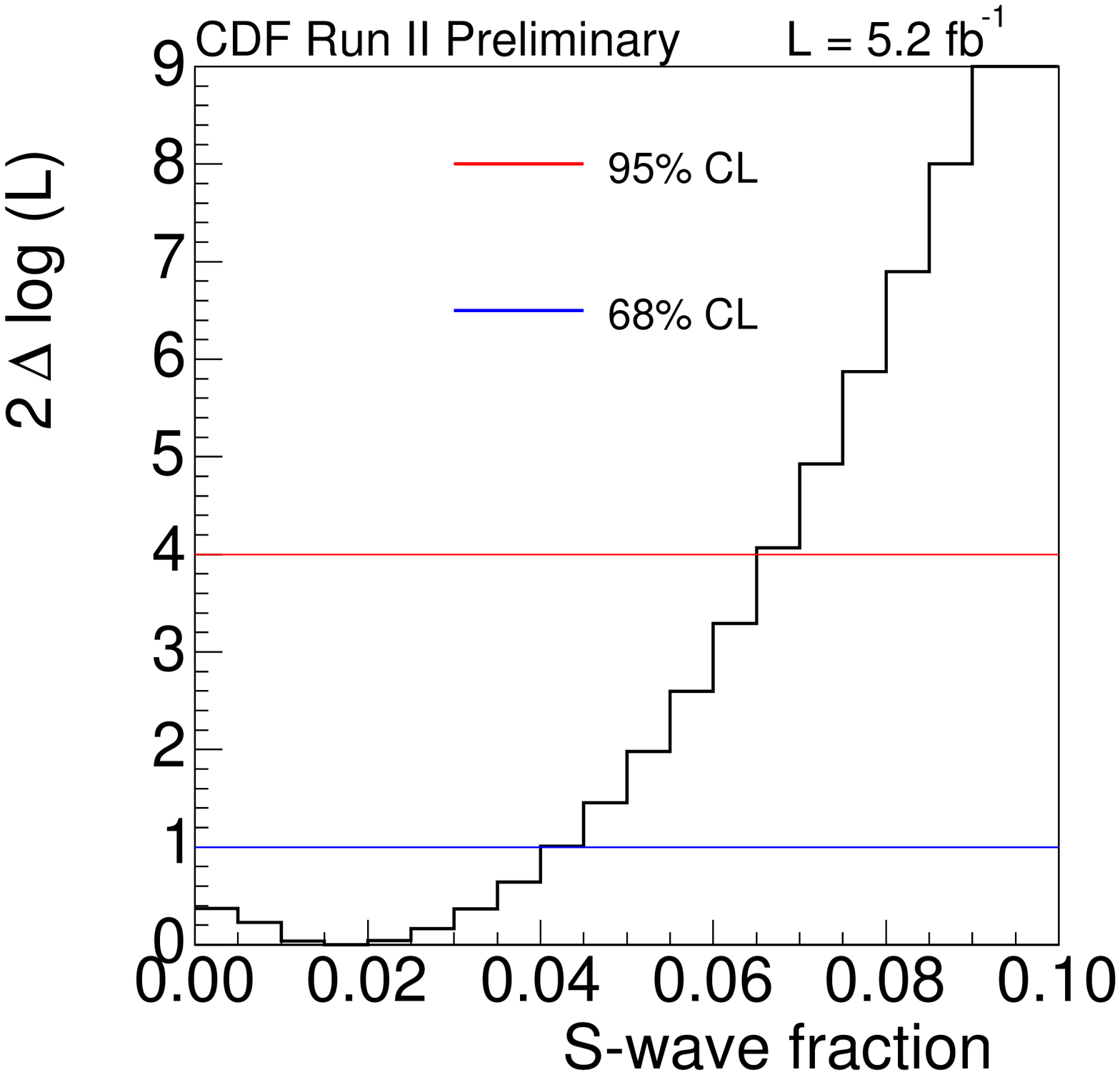}
  \caption{The contours in
$\beta_s^{\Jpsi\phi}$-$\Delta\Gamma$ plane after full
coverage adjustment (left). The likelihood profile for the fraction
of s-wave contribution (right).}
  \label{fig:result}
\end{figure}
Similar to the previous iterations, the likelihood in case of all
parameters floating still show non-Gaussian behaviour. In
order to deal with the likelihood behaviour and ensure well
defined statistical meaning of the result we construct
confidence regions in the $\beta_s^{\Jpsi\phi}$-$\Delta\Gamma$
plane, which are shown in Fig.~\ref{fig:result}. For the
standard model value of $\beta_s^{\Jpsi\phi}$ and
$\Delta\Gamma$ we obtain the p-value of 44\% which corresponds to
about 0.8 standard deviations. Minimising also over
$\Delta\Gamma$ we obtain the p-value for standard model of 31\% with
$\beta_s^{\Jpsi\phi} \in [0.02,0.52]\cup[1.08,1.55]$ at 68\%
confidence level and $\beta_s^{\Jpsi\phi} \in
[-\pi/2,-1.44]\cup[-0.13,0.68]\cup[0.89,\pi/2]$ at 95\%
confidence level. The amount of s-wave is consistent with
zero and the likelihood profile over
parameter describing its amount is shown in
Fig.~\ref{fig:result}.

In addition we also perform a fit in which we fix
$\beta_s^{\Jpsi\phi}=0$. Such configuration provides well
behaving likelihood allowing to provide point estimates of
several interesting quantities even if the result should be
interpreted in the context of standard model. From this fit
we measure
\vspace*{-0.1cm}\begin{eqnarray}
  c\tau &=& 458.6 \pm 7.6 (\mathrm{stat}) \pm 3.6 (\mathrm{syst})\, \mu \mathrm{m}, \nonumber \\
  \Delta\Gamma &=& 0.075 \pm 0.035 (\mathrm{stat}) \pm 0.01 (\mathrm{syst})\, \mathrm{ps}^{-1}, \nonumber\\
  |A_{||}|^2 &=& 0.231 \pm 0.014 (\mathrm{stat}) \pm 0.015 (\mathrm{syst}), \nonumber \\
  |A_{0}|^2 &=& 0.524 \pm 0.013 (\mathrm{stat}) \pm 0.015 (\mathrm{syst}), \nonumber \\
  \phi_\perp &=& 2.95 \pm 0.64 (\mathrm{stat}) \pm 0.07 (\mathrm{syst}). \nonumber
\end{eqnarray}\vspace*{-0.1cm}
In all cases, those are the most precise measurements of those quantities up to date.
The strong phase $\delta_{||}$ is close to the symmetry
point at $\pi$ which makes estimate of its value and
uncertainty unreliable and therefore we do not provide
result for it.

\section{Conclusions}

We presented the updated measurement of the $CP$ violating phase
$\beta_s^{\Jpsi\phi}$ in \BsJpsiphi decays from CDF experiment. Using 5.2 \invfb
of data we obtain bounds which are significantly stronger than our previous
results. The data itself are consistent with the standard model at 0.8 standard
deviation level. The improvement compared to the previous result is better than
just simple scaling by the amount of collected data. Few possible improvements
are still available in addition to collecting more data. On the data size
itself, we expect to double our dataset by the end of 2011 with ongoing discussion
for another 3 years extension to Tevatron running. 

\Acknowledgements

The author would like to thank the organisers of the CKM workshop and working
group conveners for kind invitation and for providing forum for 
discussions. Also my CDF colleagues actively pursuing this analysis deserve
acknowledgement for their hard work needed to push forward
this non-trivial but exciting analysis.

\end{document}